\documentclass[12pt]{article}

\usepackage{amssymb,amsmath}
\usepackage{hyperref}
\usepackage[mathscr]{eucal}
\usepackage{multicol}
\usepackage{amsfonts}
\usepackage{longtable}

\usepackage{graphicx}
\newcommand{\be}{\begin{equation}}
\newcommand{\ee}{\end{equation}}
\newcommand{\bea}{\begin{eqnarray}}
\newcommand{\eea}{\end{eqnarray}}
\newcommand{\s}{\star}
\newcommand{\ket}{|0\rangle}

\begin{document}

\title{\bf
\vspace{1cm}Pure Gauge Configurations\\
and  \\
Solutions to Fermionic Superstring Field Theories Equations of
Motion  }
\author{
I.Ya. Aref'eva$^1$, R.V. Gorbachev$^1$ and P.B. Medvedev$^2$\\
 \\
{\it  $^1$Steklov Mathematical Institute, Gubkin St.8, 119991, Moscow, Russia.}\\
{\it E-mail:~arefeva@mi.ras.ru,~ rgorbachev@mi.ras.ru}
\\
\\
{\it $^2$ Institute of Theoretical and Experimental Physics}
\\ {\it B. Cheremushkinskaya st. 25, 117218, Moscow, Russia.}\\
{\it E-mail:~pmedvedev@itep.ru }} \maketitle

\abstract{ Recent results on solutions to the equation of motion of
the cubic fermionic string field theory and an equivalence of
non-polynomial and cubic string field theory are discussed. To have
a possibility to deal with both $GSO(+)$ and $GSO(-)$ sectors in the
uniform way a matrix formulation for the NS fermionic SFT is used.
In constructions of analytical solutions to open string field
theories truncated pure gauge configurations parameterized by wedge
states play an essential role. The matrix form of this
parametrization for the NS fermionic SFT  is presented. Using  the
cubic open superstring field theory as an example we demonstrate
explicitly that for the large parameter of the perturbation
expansion these truncated pure gauge configurations give  divergent
contributions to the equation of motion on the subspace of the wedge
states. The perturbation expansion is cured  by adding extra terms
that are nothing but the terms necessary for the equation of motion
contracted with the solution itself to be satisfied.}

\newpage
\begin{flushright}
\textit{This work is dedicated to the memory of Aliosha
Zamolodchikov }
\end{flushright}

\section{Introduction}

It is well known that string field theories (SFT)  describe infinite
number of local fields. Just by this reason finding nontrivial
solutions to classical SFT is a rather nontrivial problem. This is a
reason why the  Schnabl construction of the tachyon solution
\cite{Sch} in the Witten open bosonic SFT \cite{W}
 attend a lot of attentions \cite{Okawa} - \cite{Ishida:2008jc}. It turns out
that the  tachyon solution  is closely related to pure gauge solutions.
More precisely, Schnabl's solution is a
regularization of a singular limit of a pure gauge configuration
\cite{Sch,Okawa}. The presence of pure gauge solutions in the bosonic SFT
is related to the Chern-Simons form of the Witten cubic action.
The Schnabl solution is distinguished by the fact that it describes a
true vacuum of SFT, i.e. a vacuum  on which the Sen conjecture is
realized. Since the pure gauge solutions do not shift the vacuum
energy the correct shift of the vacuum energy by the Schnabl
solution is rather non-trivial fact and its deep origin is still
unclear for us.

The purpose of this report is to present recent results
concerning the generalization of the Schnabl solution to the fermionic case.

It is natural to expect that a solution being a singular limit of a
pure gauge solution also  exist in the cubic super SFT (SSFT)
\cite{AMZ,PTY}. But for the superstring case there is no a priori a
reason to deal with  the Sen conjecture, since the perturbative
vacuum is stable (there is no tachyon). However a nontrivial (not
pure gauge)  solution to the SSFT equation of motion does exist
\cite{Erler}. The physical meaning of this solution is still
unclear. It may happen that it is related with a spontaneous
supersymmetry breaking (compare with \cite{AMZ2}).

There is also a non-polynomial formulation of the SSFT \cite{B}. A
solution of equation of motion for marginal deformations in the
non-polynomial SSFT has been obtained in \cite{Erler2,Okawa2}. This
construction became clear after realization an explicit relation
between solutions to the cubic and non-polynomial superstring field
theories \cite{KF}. These theories include only the $GSO(+)$ sector
of the NS string. There are also two versions of the NS fermionic
SFT that includes both $GSO(+)$ and $GSO(-)$ sectors, cubic
\cite{ABKM} and non-polynomial \cite{Ber}. Just the NS fermionic SFT
with two sectors is used to describe non-BPS branes. The Sen
conjecture has been checked by the level truncation for the
non-polynomial and cubic cases in \cite{BSZ} and \cite{ABKM},
respectively. A solution to the equation of motion of the cubic SFT
describing the NS string with both $GSO(+)$ and $GSO(-)$ sectors has
been constructed in \cite{AGM}. On this solution the Sen conjectures
takes place.

To make a construction of the solution \cite{AGM} more clear it is
useful to incorporate  a matrix version of NS fermionic SFT with
$GSO(+)$ and $GSO(-)$ \cite{ABG}. In the matrix formulation an
explicit relation between solutions to the cubic and non-polynomial
theories become more clear and it gives an explicit formula for
solutions to the non-polynomial theory  \cite{Ber} via solutions
\cite{AGM} to ABKM
 theory \cite{ABKM}.

The Schnabl solution $\Psi$ consists of two pieces and is defined by
the limit:
\begin{equation}
\Psi = \lim_{N \to \infty} \left[ \,\Psi_N(1)
 - \psi_N \, \right] \,,\,\,\,\,\Psi_N(1)=\sum_{n=0}^N \psi'_n,
\label{Psi}
\end{equation}
where the states $\psi'_n$
defined for any real $n \ge 0$
are made of the wedge state
\cite{Rastelli:2000iu, Rastelli:2001vb, Schnabl2}.

It was
shown \cite{Sch} that the string field $\Psi$ in (\ref{Psi})
solves the equation of motion of Witten's SFT
contracted with any state $C$ in the Fock space with a finite
number of string excitations.
\be \label{EOM}
\langle C,Q\Psi+\Psi\star\Psi\rangle=0.
\ee

On the other hand to  check  the Sen conjecture,
one has to use the equation of motion contracted with a solution itself.
The $\psi _N$ piece in (\ref{Psi})  is necessary
for the equation of motion contracted with the solution itself
to be satisfied \cite{Okawa,KF}.

We note that the pure gauge part of the Erler configuration  does not solve
the equation of motion contracted
 with wedge states $\psi _m$
 \be
\label{EOM}
\langle\psi _m,Q\Psi _\infty (1)+\Psi _\infty (1)\star\Psi _\infty (1)\rangle\neq 0.
\ee
It is possible to add  extra terms $\psi_N$ to $\Psi _\infty (1)$ to get a solution in the sense of \eqref{EOM}.  These are just the terms that have
been used previously to get a desirable  value of the action  \cite{Erler}.

The paper is organized as follows. In Section 2 a matrix formulation
for the NS fermionic SFT is presented. In Section 3 we contribute to
a discussion \cite{KF} of the classical equivalence of the
non-polynomial theory of Berkovits with $GSO(\pm)$ sectors
\cite{Ber} (here we refer to  this theory as the  Berkovits, Sen and
Zwiebach theory) and the cubic theory of Belov, Koshelev and two of
us \cite{ABKM}. In Section 4 perturbative parameterizations of
special pure gauge configurations are presented. These pure gauge
configurations are used in the Erler superstring field theory
solution \cite{Erler} and in the tachyon fermion solution
\cite{AGM}. We demonstrate that $\lambda=1$ limit of these pure
gauge solutions is in fact a singular point and we use a simple
prescription to cure divergences. We show that this prescription
gives the same answer as the requirement of validity of the equation
of motion contracted with the solution itself.

\section{Pure Gauge Configurations in Cubic SFT for Fermion String with $GSO(+)$ and $GSO(-)$ Sectors
in Matrix Notations}
\subsection{Cubic Fermion SFT  with $GSO(-)$ Sector
in Matrix Notations}

The action for covariant superstring field theory with $GSO(+)$ and
$GSO(-)$ sectors was proposed at \cite{ABKM}:
\begin{eqnarray}\label{9}
S[\Phi_+,\Phi_-]&=&-\frac{1}{g_0^2}\left[\frac12\langle
Y_{-2}\Phi_+,Q\Phi_+\rangle+\frac13\langle
Y_{-2}\Phi_+,\Phi_+\s\Phi_+\rangle\right.\nonumber\\
&&+\left.\frac12\langle Y_{-2}\Phi_-,Q\Phi_-\rangle-\langle
Y_{-2}\Phi_+,\Phi_-\s\Phi_-\rangle\right].
\end{eqnarray}
The equations of motion read ($\star$ stands for Witten's string field
product)
\begin{eqnarray}\label{eq_mov}
 Q\Phi_++\Phi_+\star\Phi_+-\Phi_-\star\Phi_-&=&0\label{eom},\\
 Q\Phi_-+\Phi_+\star\Phi_--\Phi_-\star\Phi_+&=&0\label{eom11}.
\end{eqnarray}
The string fields $\Phi_+$ and $\Phi_-$ have definite and opposite
Grassman parity, to be fixed below. The parity $|\Phi|$ leads to the
Leibniz rule
\begin{equation}
Q(\Phi\star\Psi)=Q\Phi\star\Psi+(-)^{|\Phi|}\Phi\star Q\Psi.
\end{equation}

It is useful to introduce matrix notations \cite{ABG} by tensoric
string fields and operators with appropriate $2\times2$ matrices. In
this notations the action (\ref{9}) reads
\begin{equation}\label{10}
S[\widehat{\Phi}]=-\frac{1}{g_0^2}\left[\frac12\langle\widehat{Y}_{-2}\widehat{\Phi},\widehat{Q}\widehat{\Phi}\rangle
+\frac13\langle
\widehat{Y}_{-2}\widehat{\Phi},\widehat{\Phi}\s\widehat{\Phi}\rangle\right],
\end{equation}
the string field $\widehat{\Phi}$ is given by \cite{ABG}
\begin{equation}\label{5}
\widehat{\Phi}=\Phi_+\otimes\sigma_3+\Phi_-\otimes i\sigma_2,
\end{equation}
and
\begin{equation}\label{6}
\widehat{Q}=Q\otimes\sigma_3,\quad
\widehat{Y}_{-2}=Y_{-2}\otimes\sigma_3,
\end{equation}
$\sigma_i$ are Pauli matrices, $Q$ is the BRST charge and $Y_{-2}$
is a double-step picture changing operator.

The parity assignment and $\sigma_i$ algebra lead to the Leibnitz
rule
\begin{equation}\label{L}
\widehat{Q}(\widehat{\Phi}\star\widehat{\Psi})=
(\widehat{Q}\widehat{\Phi})\star\widehat{\Psi}+(-)^{|\widehat{\Phi}|}\widehat{\Phi}\star(\widehat{Q}\widehat{\Psi}),
\end{equation}
where
\be
|\widehat{\Phi}|\equiv|\Phi_+|.
\ee
The action is only nonvanishing
for a string field   of degree 1.  We also have
\begin{eqnarray}
|\widehat\Phi\star\widehat\Psi|&=&|\widehat\Phi|+|\widehat\Psi|,\\
|\widehat Q\widehat\Phi|&=&1+|\widehat\Phi|,\\
|\widehat Q|&=&1,\\
|\widehat \Phi|&=&1.
\end{eqnarray}
The equations of motion (\ref{eq_mov}) in the matrix notations read
\begin{equation}\label{7}
\widehat{Q}\widehat{\Phi}+\widehat{\Phi}\star\widehat{\Phi}=0.
\end{equation}
\subsection{Pure Gauge Solutions to Equation of Motion}
Pure gauge
solutions to (\ref{7})  have the form
\begin{equation}\label{8}
\widehat{\Phi}=\widehat{\Omega}^{-1}\star\widehat{Q}\widehat{\Omega}=-\widehat{Q}\widehat{\Omega}\star\widehat{\Omega}^{-1}
\end{equation}
for $\widehat{\Phi}$ to be odd $\widehat{\Omega}$ has to be even
$|\widehat{\Omega}|=0$ . In component we have:
\begin{equation}
\widehat{\Omega}=\Omega_+\otimes I+\Omega_-\otimes\sigma_1,
\end{equation}
$\Omega_+$ and $\Omega_-$ belong to $GSO(\pm)$ sectors.

It is obvious that two pure gauge configurations are related via a gauge transformation,
\bea
\widehat{\Phi}_1&=&\widehat{\Omega}_1^{-1}\widehat{Q}\widehat{\Omega}_1,\\
\widehat{\Phi}_2&=&
\widehat{\Omega}^{-1}_2\widehat{Q}\widehat{\Omega}_2,
\eea
\bea
\widehat{\Phi}_2&=&\widehat{\Omega}^{-1}(\widehat{\Phi}_1+\widehat{Q})\widehat{\Omega},\\
\widehat{\Omega}&=&\widehat{\Omega}_1^{-1}\widehat{\Omega}_2.
\eea

A pure gauge solution in the $GSO(+)$ sector
\be
\label{Phi}
\Phi_+=\Omega^{-1}_1Q\Omega_1.
\ee
can be consider as a special pure gauge solution in the matrix case
\begin{equation}\label{phi_gauge}
\widehat{\Phi}_+=\Omega^{-1}_1Q\Omega_1\otimes\sigma_3.
\end{equation}
This configuration is gauge equivalent to a given matrix pure gauge
configuration $\,\,\,\,\,\,$  $\,\,\,\,\,\,$ $\widehat{\Phi}=\widehat{\Omega}^{-1}\widehat{Q}\widehat{\Omega}$,
i.e.
\begin{equation}
\label{phi_gauge}\widehat{\Omega}^{-1}\widehat{Q}\widehat{\Omega}=\widehat{\Omega}^{-1}_2(\Omega^{-1}_1Q\Omega_1+Q)\otimes\sigma_3\widehat{\Omega}_2.
\end{equation}
Indeed, from \eqref{phi_gauge} we get
\be
\Omega_{+2}=\Omega^{-1}_{1}\Omega_{+},\quad\Omega_{-2}=\Omega_{1}^{-1}\Omega_{-}.\ee

\subsection{Perturbative Expansion in Matrix Notations}
In this section we are going to find a solution of the equation of
motion (\ref{7}). The solution we will find as a series in some
parameter $\lambda$ i.e. let us suppose $\widehat{\Phi}$ to be a
series in some $\lambda$
\begin{equation}
\widehat{\Phi}^\lambda=\sum_{n=0}^\infty\lambda^{n+1}\widehat{\phi}_n,
\end{equation}
and put this expansion in the equation of motion (\ref{7}). In the
first order in $\lambda$ we have
\begin{equation}\label{40}
\widehat{Q}\widehat{\phi}_0=0.
\end{equation}
We choose a solution to (\ref{40}) as
\begin{equation}\label{41}
\widehat{\phi}_0=\widehat{Q}\widehat{\phi},
\end{equation}
where
\begin{equation}
\widehat{\phi}=\phi_+\otimes I+\phi_-\otimes\sigma_1,
\end{equation}
$\phi_+$ and $\phi_-$ are components of the gauge field
$\widehat{\phi}$ and they belong to $GSO(+)$ and $GSO(-)$ sectors
respectively. The Grassman parities of $\phi_+$ and $\phi_-$ are opposite.

In the second order in $\lambda$ we have
\begin{equation}\label{45}
\widehat{Q}\widehat{\phi}_1+\widehat{\phi}_0\star\widehat{\phi}_0=0.
\end{equation}
For $\widehat{\phi}_0$ in the form (\ref{41}) we get (also we used
Leibnitz rule (\ref{L}) for $\widehat{Q}$)
\begin{equation}
\widehat{Q}\widehat{\phi}_1+\widehat{Q}\widehat{\phi}\star\widehat{Q}\widehat{\phi}
=\widehat{Q}(\widehat{\phi}_1-\widehat{Q}\widehat{\phi}\star\widehat{\phi})=0,
\end{equation}
due to $|\widehat{\phi}|=0$ we get minus. The solution of equation
(\ref{45}) is
\begin{equation}
\widehat{\phi}_1=\widehat{Q}\widehat{\phi}\star\widehat{\phi}.
\end{equation}
In this scheme we get
\begin{equation}
\widehat{\phi}_n=\widehat{Q}\widehat{\phi}\star\widehat{\phi}^n,
\end{equation}
then $\widehat{\Phi}$ is
\begin{equation}\label{phi2}
\widehat{\Phi}^\lambda=\sum_{n=0}^\infty\lambda^{n+1}\widehat{Q}\widehat{\phi}\star\widehat{\phi}^n
=\lambda\widehat{Q}\widehat{\phi}\frac{1}{1-\lambda\widehat{\phi}}.
\end{equation}
The perturbative solution has the pure gauge form (\ref{8}). Indeed, let
us introduce $\widehat{\Omega}=1-\lambda\widehat{\phi}$, then
(\ref{phi2}) is
\begin{equation}
\widehat{\Phi}^\lambda=-\widehat{Q}(1-\lambda\widehat{\phi})\star(1-\lambda\widehat{\phi})^{-1}
=-\widehat{Q}\widehat{\Omega}\star\widehat{\Omega}^{-1}.
\end{equation}

This expression can be written through $\phi_{\pm}$ as
\begin{eqnarray}\label{pur}
\widehat{\Phi}^\lambda=(Q\phi_+\otimes\sigma_3+Q\phi_-\otimes
i\sigma_2)\frac{1}{(1-\lambda\phi_+)^2-\lambda^2\phi_-^2}((1-\lambda\phi_+)\otimes
I+\lambda\phi_-\otimes\sigma_1).
\end{eqnarray}
Picking out $GSO(+)$ and $GSO(-)$ sectors we get
\begin{equation}
\Phi_+^\lambda=\frac\lambda2Q(\phi_++\phi_-)\frac{1}{1-\lambda(\phi_++\phi_-)}+\frac\lambda2Q(\phi_+-\phi_-)
\frac{1}{1-\lambda(\phi_+-\phi_-)},
\end{equation}
\begin{equation}
\Phi_-^\lambda=\frac\lambda2Q(\phi_++\phi_-)\frac{1}{1-\lambda(\phi_++\phi_-)}-\frac\lambda2Q(\phi_+-\phi_-)
\frac{1}{1-\lambda(\phi_+-\phi_-)}.
\end{equation}
This result is agree with \cite{AGM}.
\section{Equivalence of BSZ and ABKM Theories}

The action for the cubic $NS$ string theory with $GSO(-)$ sector is
presented in Section 2. In the non-polynomial theory the $GSO(-)$
sector can be added by the following way \cite{Ber}. The field is an
element of $2\times2$ matrix of the form
\begin{equation}\label{36}
\widehat{G}=G_+\otimes I+G_-\otimes\sigma_1.
\end{equation}
An equation of motion has the following form
\begin{equation}\label{berk2}
\widehat{\eta}_0(\widehat{G}^{-1}\widehat{Q}\widehat{G})=0,
\end{equation}
where $\widehat{\eta}_0\equiv\eta\otimes\sigma_3$.

Let $\mathfrak{A}$ be a set of matrix solutions of equation of
motion (\ref{7}) and $\mathfrak{B}$ is set of solutions
(\ref{berk2}).

Let us define a map $g$ of $\mathfrak{B}$ to $\mathfrak{A}$  \cite{KF}
\begin{equation}\label{psi2}
g:\widehat{G}\to\widehat{\Psi}\equiv
g(\widehat{G})=\widehat{G}^{-1}\widehat{Q}\widehat{G}.
\end{equation}
This map is correctly defined due to (\ref{berk2}) ($\widehat\Psi$ is in the small Hilbert space \cite{FMS}).

In order to $\widehat{\Psi}=g(\widehat{G})$ be a solution of
equation of motion (\ref{7}) it is necessary and sufficient  to
implement
 the Leibnitz rule for operator $\widehat{Q}$ (\ref{L}). Let us note
 that $G_+$ is even and $G_-$ is odd, i.e. $G_+$ and $G_-$ have the
 different parities.

Let us define a map $h$ of $\mathfrak{A}$ in $\mathfrak{B}$ as
\cite{KF}
\begin{equation}\label{otobr}
h:\widehat{\Psi}\to\widehat{G}\equiv h(\widehat{\Psi})=e^{\widehat{P}\widehat{\Psi}}.
\end{equation}
here $\widehat{P}\equiv P\otimes\sigma_3$, where $P$ is
nilpotent operator with respect to $\star$ defined in \cite{KF}
\begin{eqnarray}
\label{x25} (P\Psi_1)\star(P\Psi_2)=0,
\end{eqnarray}
and its anticommutator with $Q$ is the identity
\begin{eqnarray}\label{25}
\{Q,P(z)\}=1.
\end{eqnarray}

Since $\widehat{P}^2=0$ we have
\begin{equation}
e^{\widehat{P}\widehat{\Psi}}=1+\widehat{P}\widehat{\Psi},
\end{equation}
here $1$ is an identity state $|I\rangle\otimes I$ with respect
to $\star$.

In the components (\ref{otobr}) reads
\begin{equation}
G_+=1+P\Psi_+=e^{P\Psi_+},\quad G_-=P\Psi_-.
\end{equation}

The maps $g$ and $h$ are connected nontrivially. Let us consider a
composition $g\circ h$:
\begin{eqnarray}
\widetilde{\widehat{\Psi}}&=&(g\circ
h)(\widehat{\Psi})=g(h(\widehat{\Psi}))=(1-\widehat{P}\widehat{\Psi})\widehat{Q}(1+\widehat{P}\widehat{\Psi})
=(1-\widehat{P}\widehat{\Psi})\widehat{Q}\widehat{P}\widehat{\Psi}\\
&=&(1-\widehat{P}\widehat{\Psi})(1-\widehat{P}\widehat{Q})\widehat{\Psi}
=(1-\widehat{P}\widehat{Q}-\widehat{P}\widehat{\Psi})\widehat{\Psi}
=\widehat{\Psi}-\widehat{P}(\widehat{Q}\widehat{\Psi}+\widehat{\Psi}^2)=\widehat{\Psi}\nonumber,
\end{eqnarray}
here, we used (\ref{25}), then we used the equation of motion for
$\widehat{\Psi}$ and the nilpotency of $\widehat{P}$  under the star
product (\ref{x25}).  So we have proved that $g\circ h=Id$ and
$g(\mathfrak{B})=\mathfrak{A}$ i.e. an arbitrary classical solution
in cubic theory can be represent in pure-gauge form.

Now let us consider a composition $h\circ g$:
\begin{eqnarray}\label{G1}
\widetilde{\widehat{G}}&=&(h\circ
g)(\widehat{G})=h(g(\widehat{G}))=e^{\widehat{P}\widehat{G}^{-1}\widehat{Q}\widehat{G}}
=1+\widehat{P}\widehat{G}^{-1}\widehat{Q}\widehat{G}=1-\widehat{P}\widehat{Q}\widehat{G}^{-1}\cdot
\widehat{G}\nonumber\\
&=&1-(1-\widehat{Q}\widehat{P})\widehat{G}^{-1}\cdot \widehat{G}
=1-1+\widehat{Q}\widehat{P}\widehat{G}^{-1}\cdot
\widehat{G}=\widehat{Q}\widehat{P}\widehat{G}^{-1}\cdot \widehat{G}.
\end{eqnarray}

For an arbitrary $\widehat{G}\in \mathfrak{B}$ introduce the
following parametrization \cite{KF}
\begin{equation}\label{param}
\widehat{G}=\frac{1}{1-\widehat{\Phi}}.
\end{equation}

\begin{figure}[h!]
    \centering
    \includegraphics[width=15cm]{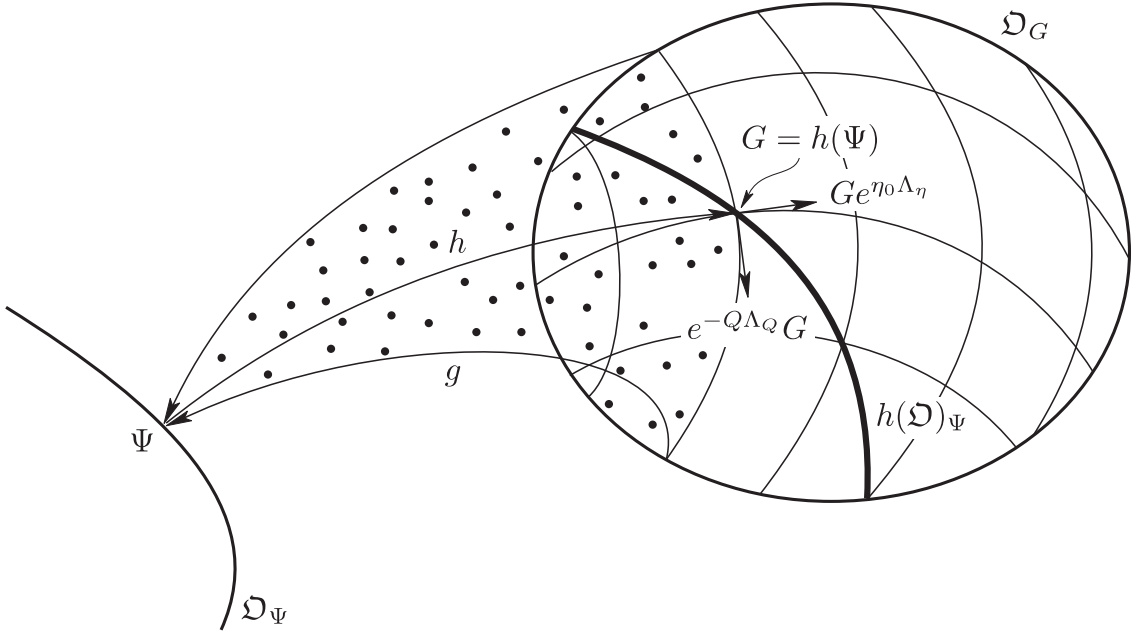}
  \caption{Maps $h$ and $g$. Here hats are omitted for simplicity. }
  \label{large-cc}
\end{figure}

The element $\widehat{\Psi}=g(\widehat{G})\in \mathfrak{A}$ takes
the form
\begin{equation}
\widehat{\Psi}=\widehat{G}^{-1}\widehat{Q}\widehat{G}=-\widehat{Q}\widehat{G}^{-1}\widehat{G}
=\widehat{Q}\widehat{\Phi}\frac{1}{1-\widehat{\Phi}}.
\end{equation}
Here it is used that $\widehat{G}$ is even, the Leibnitz rule is
used, at the same time in was important, that the parities of $G_+$
and $G_-$ are opposite and $\sigma_2I=I\sigma_2$,
$\sigma_3\sigma_1=-\sigma_1\sigma_3$. Also we used that $P$ changes
the parity of field.

Then we use the parametrization (\ref{param}) for (\ref{G1})
\begin{equation}\label{99}
\widetilde{\widehat{G}}=\widehat{Q}\widehat{P}\widehat{G}^{-1}\cdot
\widehat{G}=\widehat{Q}\widehat{P}(1-\widehat{\Phi})\cdot\frac{1}{1-\widehat{\Phi}}=
\frac{1}{1-\widehat{\Phi}}-\widehat{Q}\widehat{P}\widehat{\Phi}\frac{1}{1-\widehat{\Phi}}
=(1-\widehat{Q}(\widehat{P}\widehat{\Phi}))\widehat{G},
\end{equation}
here we use
\begin{equation}
\widehat{Q}\widehat{P}I=I.
\end{equation}
Let us rewrite (\ref{99}) as
\begin{equation}\label{88}
\widetilde{\widehat{G}}=e^{-\widehat{Q}(\widehat{P}\widehat{\Phi})}\widehat{G}.
\end{equation}
It is the gauge transformation
\begin{equation}
\widetilde{\widehat{G}}
=e^{-\widehat{Q}\widehat{\Lambda}_{\widehat{Q}}}\widehat{G}e^{\widehat{\eta}_0\widehat{\Lambda}_{\widehat{\eta}}},
\end{equation}
 with a gauge parameter
$\widehat{\Lambda}_{\widehat{Q}}=\widehat{P}\widehat{\Phi}$,
$\widehat{\Lambda}_{\widehat{\eta}}=0$.

So $(h\circ g)(\widehat{G})$ belongs to a gauge orbit
$\mathfrak{O}_{\widehat{G}} =\{\widehat{\widetilde{G}}:
\widehat{\widetilde{G}}=e^{-\widehat{Q}\widehat{\Lambda}_{\widehat{Q}}}
\widehat{G}\}$ of the initial field $\widehat{G}$. In the components
(\ref{88}) reads
\begin{eqnarray}
\widetilde{G}_+&=&e^{-Q\Lambda_+}G_+-Q\Lambda_-G_-,\nonumber\\
\widetilde{G}_-&=&e^{-Q\Lambda_+}G_--Q\Lambda_-G_+,
\end{eqnarray}
\begin{equation}
\widehat{\Lambda}_Q=\Lambda_+\otimes\sigma_3+\Lambda_-\otimes
i\sigma_2,
\end{equation}
where $\Lambda_+=P\Phi_+,\quad \Lambda_-=P\Phi_-$.

In the terms of gauge orbits the maps $g$ and $h$ can be describe
more clearly.

Let $\widehat{\Psi}$ be an arbitrary field of $\mathfrak{A}$ and
$\widehat{G}=h(\widehat{\Psi})$. Let us consider an image of the
orbit $\mathfrak{O}_{\widehat{\Psi}}
=\{\widetilde{\widehat{\Psi}}:\widetilde{\widehat{\Psi}}=e^{-\widehat{\Lambda}}(\widehat{\Psi}+\widehat{Q})e^{\widehat{\Lambda}}\}$
 by the map
$h$: $h(\mathfrak{O}_{\widehat{\Psi}})=
\{\widetilde{\widehat{G}}:\widetilde{\widehat{G}}=h(\widetilde{\widehat{\Psi}})\}
$. The direct calculation gives:
\begin{eqnarray}\label{psi}
\widetilde{\widehat{G}}&=&1+\widehat{P}\widetilde{\widehat{\Psi}}
=1+\widehat{P}(e^{-\widehat{\Lambda}}(\widehat{\Psi}+\widehat{Q})e^{\widehat{\Lambda}})
=1+\widehat{P}(-\widehat{Q}e^{-\widehat{\Lambda}}+e^{-\widehat{\Lambda}}\widehat{\Psi})e^{\widehat{\Lambda}}
\nonumber\\&=&(\widehat{Q}(Pe^{-\widehat{\Lambda}})+\widehat{P}e^{-\widehat{\Lambda}}\widehat{\Psi})e^{\widehat{\Lambda}}
=\widehat{Q}(\widehat{P}e^{-\widehat{\Lambda}})(1+\widehat{P}\widehat{\Psi})e^{\widehat{\Lambda}}
=\widehat{Q}(\widehat{P}e^{-\widehat{Q}\widehat{P}\widehat{\Lambda}})\widehat{G}e^{\widehat{\Lambda}}\nonumber\\
&=&
e^{-\widehat{Q}\widehat{P}\widehat{\Lambda}}\widehat{G}e^{\widehat{\Lambda}}
=e^{-\widehat{Q}\widehat{P}\widehat{\Lambda}}\widehat{G}e^{\widehat{\eta}_0\widehat{\xi}\widehat{\Lambda}},
\end{eqnarray}
i.e. $ h(\mathfrak{O}_{\widehat{\Psi}} )$ is suborbit  of the field
$\widehat{G}=h(\widehat{\Psi})$, due to a special choose of gauge
parameter $\widehat{\Lambda}_{\widehat{Q}},~\widehat{\Lambda}_{\widehat{\eta}}$
or $h(\mathfrak{O}_{\widehat{\Psi}} )\subset
\mathfrak{O}_{\widehat{G}} $.

Let $\widehat{G}$ be an arbitrary field of $\mathfrak{B}$ and $\widehat{\Psi}
=g(\widehat{G})$. Let us consider an image of the orbit
$\mathfrak{O}_{\widehat{G}}$ by the map $g$: $
g(\mathfrak{O}_{\widehat{G}})=\{\widetilde{\widehat{\Psi}}=g(\widetilde{\widehat{G}}):
\widetilde{\widehat{G}}\in \mathfrak{O}_{\widehat{G}} \}$:
\begin{eqnarray}\label{g}
\widetilde{\widehat{\Psi}}&=&\widetilde{\widehat{G}}^{-1}\widehat{Q}\widetilde{\widehat{G}}
=e^{-\widehat{\eta}_0\widehat{\Lambda}_{\widehat{\eta}}}\widehat{G}^{-1}e^{\widehat{Q}\widehat{\Lambda}_{{\widehat{Q}}}}
\widehat{Q}(e^{-\widehat{Q}\widehat{\Lambda}_{\widehat{Q}}}\widehat{G}e^{\widehat{\eta}_0\widehat{\Lambda}_{\widehat{\eta}}})\nonumber\\
&=&e^{-\widehat{\eta}_0\widehat{\Lambda}_{\widehat{\eta}}}\widehat{G}^{-1}
((\widehat{Q}\widehat{G})e^{\widehat{\eta}_0\widehat{\Lambda}_{\widehat{\eta}}}
+\widehat{G}\widehat{Q}e^{\widehat{\eta}_0\widehat{\Lambda}_{\widehat{\eta}}})
=e^{-\widehat{\eta}_0\widehat{\Lambda}_{\widehat{\eta}}}
(\widehat{\Psi}+\widehat{Q})e^{\widehat{\eta}_0\widehat{\Lambda}_{\widehat{\eta}}},
\end{eqnarray}
since $\widehat{\Lambda}_{\widehat{\eta}}$ is arbitrary, then
$g(\mathfrak{O}_{\widehat{G}})= \mathfrak{O}_{\widehat{\Psi}} $.
Note that, if $h(\widehat{\Psi} ')\in
\mathfrak{O}_{h(\widehat{\Psi})}$, then $\widehat{\Psi} '\in
\mathfrak{O}_{\widehat{\Psi}} $. Indeed, by virtue of $g\circ h=Id$ it
is possible to rewrite $\widehat{\Psi}' =g(h(\widehat{\Psi} '))$,
and since $g(\mathfrak{O}_{\widehat{G}})=
\mathfrak{O}_{\widehat{\Psi}} $, then $h(\widehat{\Psi} ')\in
\mathfrak{O}_{h(\widehat{\Psi})}$.

So we can see that the maps $g$ and $h$ could be constrict to the
maps orbits:
 $$
h:\mathfrak{O}_{\widehat{\Psi}}\to\mathfrak{O}_{\widehat{G}},~~g:\mathfrak{O}_{\widehat{G}}\to\mathfrak{O}_{\widehat{\Psi}}
 $$
At the same time the image $\mathfrak{O}_{\widehat{\Psi}}$ in
$\mathfrak{O}_{\widehat{G}}$ is suborbit (\ref{psi}). The image
$\mathfrak{O}_{\widehat{G}} =\{\widetilde{\widehat{G}}:
\widetilde{\widehat{G}}=e^{-\widehat{Q}\widehat{\Lambda}_{\widehat{Q}}}
\widehat{G} e^{\widehat{\eta}_0 \widehat{\Lambda}_{\widehat{\eta}}}
\}$ is all orbit $\mathfrak{O}_{\widehat{\Psi}}$. All elements
$\mathfrak{O}_{\widehat{G}}$ with different
$\widehat{\Lambda}_{\widehat{Q}}$ are mapped in one element
$\mathfrak{O}_{\widehat{\Psi}}$ (see (\ref{g})). Bounded on
$h(\mathfrak{O}_{\widehat{\Psi}})$ mapping $g$ becomes  invertible:
$h\circ g|_{h(\mathfrak{O}_{\widehat{\Psi})}}=Id$. The composition
$h\circ g$ gives in the orbit $\mathfrak{O}_{\widehat{G}}$ a special
section (\ref{G1}). See figure 1.

\section{Perturbative  Expansion of Pure Gauge Configurations}
\subsection{Initial data and Formal Perturbative  Expansion in Components}
Here we choose $\phi_+$ and $\phi_-$ in the following form
\cite{AGM}
\begin{eqnarray}
\label{phi}
\phi_+ &=&B^L_1c_1|0\rangle,\\
\label{psi}
\phi_-&=&B^L_1\gamma_{\frac12}|0\rangle.
 \end{eqnarray}
Then $\Phi_+^\lambda$ and $\Phi_-^\lambda$ will have the form
\begin{eqnarray}
\label{final-Phi}
\Phi_+^\lambda&=&\sum_{n=0}^\infty\lambda^{n+1}\phi^\prime_{n},\\
\phi^\prime_{0}&=&\left(-K_1^Rc_1-B^R_1(c_0c_1+\gamma
^2_{1/2})\right)|0\rangle
\label{final-Phi-0},\\
\phi_{n}^\prime&=& c_1|0\rangle\star |n\rangle\star
K^L_1B^L_1c_1|0\rangle+\gamma_{\frac12}|0\rangle\star |n\rangle\star
K^L_1B^L_1\gamma_{\frac12}|0\rangle,\,\,\, n>0, \label{final-Phi-n}
\end{eqnarray}
\begin{eqnarray}\label{final-Psi}
\Phi_-^\lambda&=&\sum_{n=0}^\infty\lambda^{n+1}\psi^\prime_{n},\\
\psi^\prime_{0}&=&\left(-K^R_1\gamma_{\frac12}+B^R_1(c_1\gamma_{-\frac12}-
\frac12c_0\gamma_{\frac12})\right)|0\rangle
\label{final-Psi-0},\\
\psi^\prime_{n}&=&\gamma_{\frac12}|0\rangle\star |n\rangle\star
K^L_1B^L_1c_1 |0\rangle+c_1|0\rangle\star |n\rangle\star
K^L_1B^L_1\gamma_{\frac12}|0\rangle\,,\,\, n>0. \label{final-Psi-n}
\end{eqnarray}

\subsection{$\lambda=1$ limit}
In this section we examine $\lambda=1$ limit of the pure gauge
solutions (\ref{pur}). It is known that this is a  singular point
for the pure gauge solution \cite{Sch,Okawa}.

We consider for the transparency the pure $GSO(+)$ sector and the
equation of motion for the string field $\Phi_+$  is
\begin{equation}\label{gso+}
Q\Phi_++\Phi_+\star\Phi_+=0.
\end{equation}

We start with the pure gauge solution to (\ref{gso+}) given by
formulae (\ref{final-Phi}) -- (\ref{final-Psi-n}) with $\lambda <1$
and initial date $\phi_-=0$. The explicit form of this solution is
\begin{equation}
\label{ser}
\Phi_+(\lambda)=\sum_{n=0}^\infty\lambda^{n+1}\varphi_n'+\lambda
\Gamma,\qquad |\lambda| <1,
\end{equation}
where
\begin{eqnarray}
\label{ser-1}
\Gamma&=&B^L_1\gamma^2_{1/2}\ket,\nonumber\\
\varphi^\prime_{0}&=&-\left(K_1^Rc_1+B^R_1c_0c_1\right)|0\rangle,\\
\varphi_{n}^\prime&=& c_1|0\rangle\star |n\rangle\star
K^L_1B^L_1c_1|0\rangle\,\,\, n>0.\nonumber
\end{eqnarray}

Let us take just a partial sum of the infinite series (\ref{ser})
\begin{equation}
\label{reg}
\Phi_+^{N}(\lambda)=\sum_{n=0}^{N-1}\lambda^{n+1}\varphi_n'+\lambda\Gamma,
\end{equation}
and check a validity of the
equation of motion (\ref{gso+}) in a weak sense on
the states $\varphi_m$ \footnote{Here
$\langle\langle...\rangle\rangle=\langle Y_{-2}...\rangle$}
\begin{equation}
\langle\langle\varphi_m,Q\Phi^{N}_+(\lambda)+\Phi^{N}_+(\lambda)\s\Phi^{N}_+(\lambda)\rangle\rangle,
\end{equation}
where \be \varphi_m=\frac2\pi c_1|0\rangle\star |m\rangle\star
B^L_1c_1|0\rangle.\ee

We use correlators \cite{Erler} collected in the table below
\begin{eqnarray}
\label{correlators}
\langle\langle\varphi_m,Q\varphi_n\rangle\rangle&=&-\frac{m+n+2}{\pi^2},\nonumber\\
\langle\langle\varphi_m,Q\Gamma\rangle\rangle&=&\frac{1}{\pi^2},\nonumber\\
\langle\langle\Gamma,Q\Gamma\rangle\rangle&=&0,\\
\langle\langle\varphi_k,\varphi_m\star\varphi_n\rangle\rangle&=&
0,\nonumber\\
\langle\langle\Gamma,\varphi_m\star\varphi_n\rangle\rangle&=&
\frac{m+n+3}{2\pi^2},\nonumber\\
\langle\langle\Gamma,\Gamma\star\varphi_n\rangle\rangle&=&
0,\nonumber\\
\langle\langle\Gamma,\Gamma\star\Gamma\rangle\rangle&=&
0.\nonumber
\end{eqnarray}

We get
\begin{equation}
\label{uon} \langle\langle
\varphi_m,Q\Phi_+^N(\lambda)+\Phi_+^N(\lambda)\star\Phi_+^N(\lambda)\rangle\rangle=\frac{\lambda^{N+1}}{\pi^2}.
\end{equation}

Taking the limit $N\to\infty$ for $\lambda<1$ we have for an
arbitrary $m$
\begin{equation}
\langle\langle\varphi_m,Q\Phi_+(\lambda)+\Phi_+(\lambda)\s\Phi_+(\lambda)\rangle\rangle=0,
\end{equation}
in other words for $\lambda <1$  the field $\Phi_+(\lambda)$ solves
the E.O.M. when contracted with  states from the subspase ${\cal L}(\{\varphi_m\})$ spanned by
$\varphi_m$. This fact is natural  for the solution obtained by the iteration
procedure. It is interesting to note that if
we consider the validity of the equation of motion on the subspace spanned by
$\varphi^\prime_m$ we get that on this subspace the equation of motion
are satisfied for any $\lambda$

\begin{equation}
\langle\langle\varphi^\prime_m,Q\Phi_+(\lambda)+\Phi_+(\lambda)\s\Phi_+(\lambda)\rangle\rangle=0.
\end{equation}

From equation (\ref{uon}) one sees  that for $\lambda=1$ the string
field  $\Phi_+\equiv\Phi_+(1)$ does not solve the equation of motion
(\ref{gso+}) in the week sense on ${\cal L}(\{\varphi_m\})$ \be
\label{SB-eom}
\langle\langle\varphi_m,Q\Phi_+(1)+\Phi_+(1)\s\Phi_{+}(1)\rangle\rangle =\frac{1}{\pi^2}.
\ee

Let us remind that in the case of boson string to ensure the equation of motion  in the sense
(\ref{SB-eom})  extra terms have been added to $\Phi^N_{bos}$ \cite{AGGKMM} and
these extra terms provide the validity of the Sen conjecture
\cite{Sch,Okawa}.

Following Erler \cite{Erler} we can try to add to $\Phi_+^{N}\equiv \sum_{n=0}^{N-1}\varphi_n'+\Gamma$ two extra terms
\bea
\label{reg2}
\Phi_+^{N}(c_1,c_2)=\Phi_+^{N}+c_1\varphi_N+c_2\varphi_N'
\eea
and find $c_1$ and $c_2$ from a requirement of the validity of the equation of motion in the weak sense,
\begin{equation}
\langle\langle
\varphi_m,Q\Phi^{N}_+(c_1,c_2)+\Phi^{N}_+(c_1,c_2)\star\Phi^{N}_+(c_1,c_2)\rangle\rangle=0.
\end{equation}
Simple calculations based on (\ref{correlators}) show that $c_1=-1$ and $c_2$ is arbitrary.
Indeed,
\begin{eqnarray}
\langle\langle\varphi_m,Q\Phi_+^{N}(c_1,c_2)\rangle\rangle&=&-\frac{N-1}{\pi^2}-c_1\frac{m+N+2}{\pi^2}-c_2\frac{1}{\pi^2},\nonumber\\
\langle\langle\varphi_m,\Phi_+^{N}(c_1,c_2)\star\Phi_+^{N}(c_1,c_2)\rangle\rangle&=&
\frac{N}{\pi^2}+c_1\frac{m+N+3}{\pi^2}+c_2\frac{1}{\pi^2}
\end{eqnarray}
and we see that
\begin{eqnarray}
&\,&\langle\langle\varphi_m,Q\Phi_+^{N}(c_1,c_2)+\Phi_+^{N}(c_1,c_2)\star\Phi_+^{N}(c_1,c_2)\rangle\rangle\nonumber\\
&=&-\frac{N-1}{\pi^2}-c_1\frac{m+N+2}{\pi^2}-c_2\frac{1}{\pi^2}+\frac{N}{\pi^2}+c_1\frac{m+N+3}{\pi^2}+c_2\frac{1}{\pi^2}\nonumber\\
&=&\frac{1}{\pi^2}+c_1\frac{1}{\pi^2}
\end{eqnarray}
is equal to zero for $c_1=-1$.

 Let us add to our subspace ${\cal L}(\{\varphi_m\})$ a vector $\Gamma$
and consider the requirement of the validity of the equation of
motion also on this vector \be \langle\langle
\Gamma,Q\Phi^{N}_+(-1,c_2)+\Phi^{N}_+(-1,c_2)\star\Phi^{N}_+(-1,c_2)\rangle\rangle=0.
\ee We have
\begin{equation}
\label{Ga-eq}
\langle\langle
\Gamma,Q\Phi_+^{N}(-1,c_2)+\Phi_+^{N}(-1,c_2)\star\Phi_+^{N}(-1,c_2)\rangle\rangle=-\frac{1}{\pi^2}+\frac{3}{2\pi^2}-c_2\frac{1}{\pi^2}.
\end{equation}
and we see that the L.H.S. of (\ref{Ga-eq}) is zero for $c_2=1/2$.
So
\begin{equation}
\Phi^N_+(-1,1/2)=\sum_{n=0}^{N-1}\varphi'_n+\Gamma-\varphi_N+\frac12\varphi_N'.
\end{equation}

It is interesting to note that $c_1=-1, c_2=1/2$ provide the validity of the equation of motion
being contracting with $\Phi_+^N(-1,1/2)$
\begin{equation}
\langle\langle
\Phi_+^N(-1,1/2),Q\Phi_+^{N}(-1,1/2)\,\,+\,\,\Phi_+^{N}(-1,1/2)\star\Phi_+^{N}(-1,1/2)\rangle\rangle=0.
\end{equation}

Therefore, we see that just the requirement of the validity of E.O.M. "terms by terms" at the point $\lambda=1$
forces one to add two extra terms to $\Phi^N_+$. A necessity of these extra terms
have been advocated in \cite{Erler} to provide   the Sen conjecture.

\section{Conclusion}
In this article  a singular limit of the pure gauge solution is
discussed. We propose a simple recept to deal with a singularity
problem and  on the example of cubic SSFT show that it gives the same answer as the requirement
 to get a desirable  value of the action  \cite{Erler}  (see the discussion of the same question for the case
with $GSO(-)$ sector in \cite{AGGKMM}
)

The equivalence of the solutions of the equation of motion in the
cubic fermionic string field theory \cite{ABKM} and that  of the
non-polynomial fermionic string field theory \cite{Ber} including
the $GSO(-)$ sectors is discussed using the matrix representations
of both theories. However  the singularity problem recall  that  a
formal gauge equivalence
 of two theories needs  a rather delicate studies.

The work is supported in part by RFBR grant 08-01-00798 and
NS-795.2008.1.

\end{document}